\documentclass{emulateapj}
\usepackage{amsmath}
\usepackage{graphicx}
\usepackage{amssymb}

\newcommand{\farcsec}{\mbox{\ensuremath{.\!^{\prime\prime}}}}
\newcommand{\farcmin}{\mbox{\ensuremath{.\!^{\prime}}}}
\newcommand{\degree}{\ensuremath{^{\circ}}}
\newcommand{\cir}{Circinus X-1}

\begin{document}

\title{Parsec-Scale Bipolar X-ray Shocks Produced by Powerful Jets
  from the Neutron Star Circinus X-1}

\author{
P. H. Sell\altaffilmark{1},
S. Heinz\altaffilmark{1},
D. E. Calvelo\altaffilmark{2},
V. Tudose\altaffilmark{3,4,5},
P. Soleri\altaffilmark{6},
R. P. Fender\altaffilmark{2},
P. G. Jonker\altaffilmark{7,8,9},
N. S. Schulz\altaffilmark{10},
W. N. Brandt\altaffilmark{11},
M. A. Nowak\altaffilmark{10},
R. Wijnands\altaffilmark{12},
M. van der Klis\altaffilmark{12},
P. Casella\altaffilmark{2}
}
\altaffiltext{1}{Department of Astronomy, University of Wisconsin-Madison, Madison, WI~53706, USA}
\altaffiltext{2}{School of Physics and Astronomy, University of Southampton, Southampton SO17 1BJ, UK}
\altaffiltext{3}{Netherlands Institute for Radio Astronomy, Postbus 2, 7990 AA Dwingeloo, The Netherlands}
\altaffiltext{4}{Astronomical Institute of the Romanian Academy, Cutitul de Argint 5, RO-040557 Bucharest, Romania}
\altaffiltext{5}{Research Center for Atomic Physics and Astrophysics, Atomistilor 405, RO-077125 Bucharest, Romania}
\altaffiltext{6}{Kapteyn Astronomical Institute, University of Groningen, P.O. Box 800, 9700 AV Groningen, The Netherlands}
\altaffiltext{7}{SRON, Netherlands Institute for Space Research, Sorbonnelaan 2, 3584~CA, Utrecht, The Netherlands}
\altaffiltext{8}{Department of Astrophysics, IMAPP, Radboud University Nijmegen, P.O. Box 9010, NL-6500 GL Nijmegen, the Netherlands}
\altaffiltext{9}{Harvard--Smithsonian Center for Astrophysics, 60 Garden Street, Cambridge, MA~02138, USA}
\altaffiltext{10}{Kavli Institute for Astrophysics and Space Research, Massachusetts Institute of Technology, Cambridge, MA 02139, USA}
\altaffiltext{11}{Department of Astronomy and Astrophysics, The Pennsylvania State University, 525 Davey Laboratory, University Park, PA 16802, USA}
\altaffiltext{12}{Astronomical Institute Anton Pannekoek, University of Amsterdam, Science Park 904, 1098XH Amsterdam, The Netherlands}

\begin{abstract}

  We report the discovery of multi-scale X-ray jets from the accreting
  neutron star X-ray binary, Circinus~X-1.  The bipolar outflows show
  wide opening angles and are spatially coincident with the radio jets
  seen in new high-resolution radio images of the region.  The
  morphology of the emission regions suggests that the jets from
  Circinus~X-1 are running into a terminal shock with the interstellar
  medium, as is seen in powerful radio galaxies.  This and other 
  observations indicate that the jets have a wide opening angle, 
  suggesting that the jets are either not very well collimated or
  precessing.  We interpret the spectra from the shocks as cooled
  synchrotron emission and derive a cooling age of $\sim 1600$ yr.
  This allows us to constrain the jet power to be $3 \times
  10^{35}\,{\rm erg\,s^{-1}} \lesssim P_{\rm jet} \lesssim
  2 \times 10^{37}\,{\rm erg\,s^{-1}}$, making this one of a few microquasars
  with a direct measurement of its jet power and the only known
  microquasar that exhibits stationary large-scale X-ray emission.

\end{abstract}

\keywords{ISM: jets and outflows, X-rays: binaries, X-rays: individual
  (Circinus X-1)}

\section{Introduction} \label{section:intro}

Circinus~X-1 is an unusual X-ray binary (XRB).  \cite{stewart93} first
noticed that Circinus~X-1 resides inside of a parsec-scale radio nebula
inflated by curved jets.  More-recent radio observations have observed
jet emission on multiple scales \citep[][and references
therein]{tudose08}.  Additionally, at the peak of the 40+ year X-ray
light curve \citep[see Figure 1 of][]{parkinson03}, \citet{fender04}
detected evidence of superluminal motion from ultra-relativistic jets
within a few arcseconds of the XRB, which would make this the only
known accreting neutron star with an ultra-relativistic jet.
With its resolved, parsec-scale radio jets, Circinus~X-1 has become an
important stepping stone in our understanding of microquasars over the
past few decades.

Circinus~X-1 has been extensively studied at X-ray wavelengths.
Type-I X-ray bursts, originally observed by \cite{tennant86} and seen
again by \cite{linares10} during a recent flare, now firmly establish
that the compact object is an accreting neutron star.  Multiple {\em
  Chandra} gratings observations of the point source have revealed
variable X-ray P-Cygni line profiles that were interpreted as 
high-velocity outflows (\citealt{brandt00}; \citealt{schulz02}).
\cite{iaria08} have claimed possible detection of a highly-inclined
precessing jet (similar to SS433, e.g., \citealt{lopez06}) through
Doppler-shifted X-ray line emission.  However, this interpretation is
not unique as the emission is also consistent with simple orbital
motion of the neutron star.  Moreover, \cite{schulz08} also
analyzed that particular data set and, with systematic uncertainties
included, find no shifts.  Finally, in the longest gratings
observation of the point source (zero-order, 50~ks) taken in 2005,
\citet{heinz07b} discovered faint, diffuse X-ray emission roughly
coincident with the arcminute-scale radio jets.  However, the
observation had insufficient signal--to--noise to constrain the nature
of the emission spectroscopically.  Similar diffuse emission was
subsequently seen in a 50~ks HRC-I observation \citep{soleri09}, but
no spectral information was available with this instrument.

In this Letter, we report on a follow-up deep {\em Chandra} imaging
observation of the diffuse X-ray emission found by \citet{heinz07b}.
We outline our observations in Section 2 and present an initial analysis of
the emission in Section 3. Section 4 discusses the physical implications.
Finally, we summarize our results in Section 5.  Throughout this Letter, we
assume a distance of 7.8~kpc \citep{jonker07}, which is in the middle
of a wide range of distance estimates (4.1--11.8 kpc;
\citealt{iaria05}).

\section{Observations} \label{section:obs}

We observed Circinus~X-1 on the S3 chip of the Advanced CCD Imaging
Spectrometer (ACIS) aboard the \emph{Chandra X-ray Observatory} on
2009 May 1, in a continuous 99~ks exposure. Data were taken in timed
exposure mode and telemetered in Faint mode.  Data reduction and
analysis were completed using CIAO and Sherpa versions 4.2, XSpec
12.5.1, and ACIS Extract version 2010-02-26 (AE; \citealt{broos10}).
Compared to previous gratings and HRC-I observations, this observation
proved about an order of magnitude more sensitive to diffuse flux not
only because of the long exposure and the lack of gratings, but also
because the contaminating emission from the point source was at an
exceptionally low level (Section 3.1).

\begin{figure}[tbp]
  \centering
  \rotatebox{90}{\resizebox{3.15in}{!}{\includegraphics{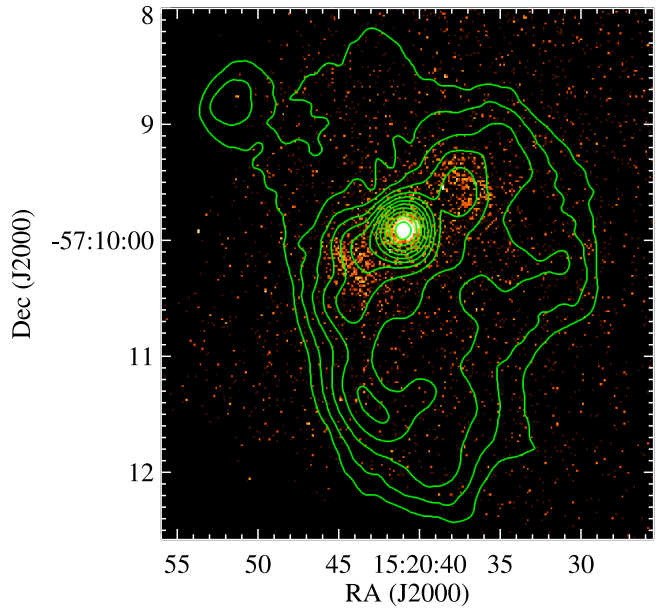}}}
  \caption{Reduced event file showing the point source and the
    bipolar ``caps'' created by the outflow. Overlaid are contours
    from the lower resolution radio image (logarithmically spaced
    from 1.9 to 25 ${\rm mJy} \ {\rm beam}^{-1}$; beam size:  
    $18 \farcsec 1 \times 16 \farcsec 5$) from \cite{tudose06}.
    \label{fig:outer_radio_overlay}}
\end{figure}

Radio comparison data were constructed from two separate observations.
Standard calibrations using the Miriad software \citep{sault95} were
applied to both sets of observations. A low angular resolution radio
image was derived from data taken on 2001 August 3 during an 11 hr run
at 1.4 GHz with the Australia Telescope Compact Array (ATCA) in 1.5A
array configuration.  This observation and the resulting images were
presented in \citet{tudose06} and are shown as contours in
Figure~\ref{fig:outer_radio_overlay}.

\begin{figure}[tbp]
  \centering
  \rotatebox{90}{\resizebox{3in}{!}{\includegraphics{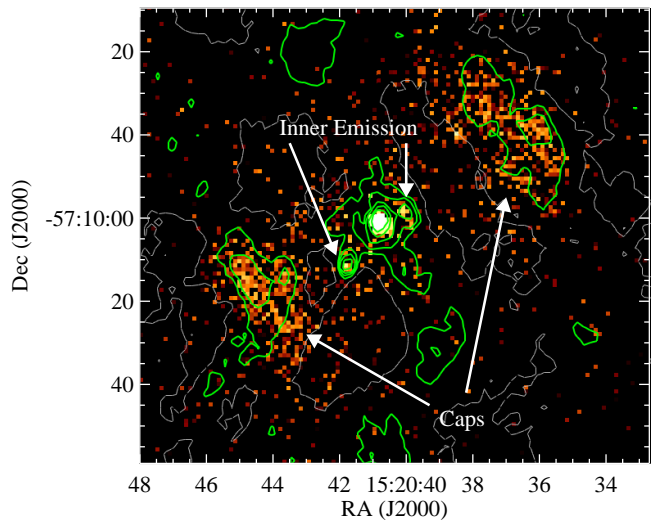}}}
  \caption{Exposure-corrected and azimuthally-smoothed,
    background-subtracted X-ray image. Overlaid are contours from the
    higher resolution radio image (levels: -1, 2, 4, 8, 16, 32 times
    rms noise of $\sim 6 \ \mu {\rm Jy} \ {\rm beam}^{-1}$; beam size:
    $3 \farcsec 02 \times 2 \farcsec 21$; D. E. Calvelo et al., in preparation).  The
    point source, inner emission, and the X-ray caps described in Sections~3.1--3.3
    are all evident here and well-matched to the radio contours.
    \label{fig:inner_radio_overlay}}
\end{figure}

A higher angular resolution radio image (Figure~
\ref{fig:inner_radio_overlay}) was created from $\sim 80$ hr of
observations spread from 2009 December 30 to 2010 January 8, using the
ATCA-Compact Array Broadband Backend in 6A configuration at 5.5 GHz
(D. E. Calvelo et al., in preparation).

\section{Analysis} \label{section:analysis}

The reduced event file of the X-ray observation is shown in
Figure~\ref{fig:outer_radio_overlay} with a contour overlay of
the low-resolution radio image of the large-scale radio nebula of
Circinus~X-1.  This image clearly shows the existence of two bright emission
regions $\sim 30 \arcsec$ from the point source.

In order to highlight the morphology of the diffuse emission closer to
the point source and to allow detailed comparison with the
high-resolution radio image, we created an exposure-corrected,
background-subtracted image (Figure~\ref{fig:inner_radio_overlay}).
Background surface brightness profiles were constructed by azimuthally
averaging over nested partial annuli in the NE and SW quadrants
(visually chosen to exclude the excess diffuse emission) and then
subtracted from the exposure-corrected image.

\begin{figure}[tbp]
  \centering
  \resizebox{!}{\columnwidth}{\includegraphics{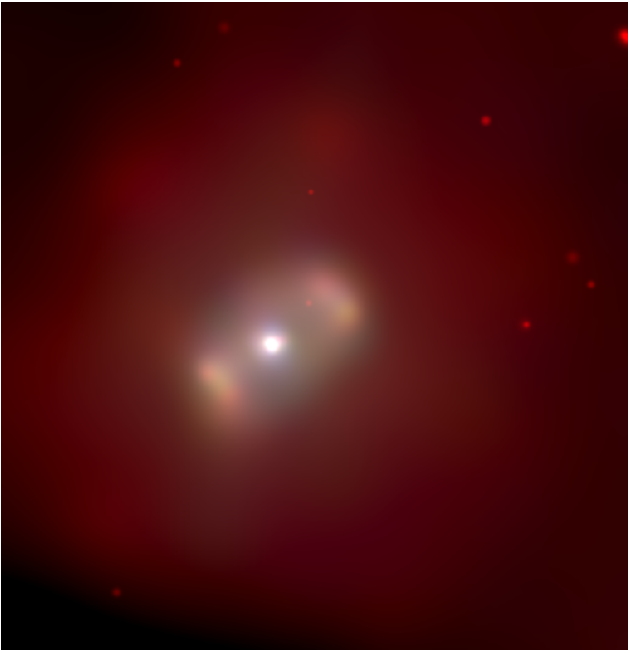}}
  \caption{Adaptively smoothed three-color image $\sim 4 \farcmin 5 \times
    4 \farcmin 5$ (red:~0.5--2.9 keV, green:~2.9--4.2 keV, and blue:~
    4.2--7.0 keV).  The large-scale dust-scattering halo from the point
    source (red) and the hard wings of the PSF (blue) are clearly
    distinct in color from the cap emission.  North is up and east is left.
    \label{fig:csmooth}}
\end{figure}

Finally, we note that on the largest scales, we find roughly axisymmetric 
soft emission from what we interpret as the dust-scattering halo, centered on the 
binary, best visible in the adaptively-smoothed image in Figure~\ref{fig:csmooth}.  
An in-depth discussion of the dust-scattering halo will be presented in a follow-up paper.

Image analysis and X-ray--radio comparisons are presented below, ordered
from smaller to larger scales.  All well-determined background/foreground 
point sources were masked and excluded from our analysis.

\subsection{The Point Source} \label{section:pt_src} 

We caught Circinus~X-1 at the lowest observed flux to date 
($F_{\rm 0.5-8 keV} \simeq 1 \times 10^{-11}\,{\rm erg\,cm^{-2}\,s^{-1}}$).
Even so, the source is still significantly piled up with a pileup fraction
of $\sim 40\%$.  We will present detailed model fits to the point
source spectrum in a follow--up paper.

\subsection{An Arcsecond Jet} \label{section:inner_emission}

Because the point source was so faint, the point spread function (PSF)
from the XRB only dominates the inner $\sim 5 \arcsec$.  This enabled us to search for
jet emission on scales of a fraction of a parsec, much smaller than
what had been possible in \cite{heinz07b} and \cite{soleri09}.  We
found two sources within $\sim 15 \arcsec$ of the point source, which
are both evident in Figure~\ref{fig:inner_radio_overlay}:

\begin{itemize}
\item[1.]{In the W-NW direction (between position angles $0^{\circ}$ and
    $+45^{\circ}$, measured counterclockwise from due W), we find a
    clear surface brightness enhancement separated from the XRB by
    $\sim 5 \arcsec$ and extending out to $\sim 8 \arcsec$.  The
    emission appears to be {\em resolved} and is coincident with a
    source in the high-resolution radio image.  No IR counterpart is
    found in either 24 $\mu$m
    MIPSGAL\footnote{http://mipsgal.ipac.caltech.edu/} or 3.6--8$\mu$m
    GLIMPSE\footnote{http://www.astro.wisc.edu/sirtf/} {\em Spitzer} images.
    Given that the source is extended and aligned with the large-scale
    radio jet and with the X-ray caps reported below, we interpret
    this source as genuine jet emission.}
\item[2.]{In the SE direction (at position angle $230 \degree$), we detect
    a bright, {\em unresolved} point source at $\sim 13 \arcsec$ from
    the XRB.  This emission is coincident with a point-like radio
    source. It is spatially coincident with a very bright IR point
    source, with no optical counterpart detected in an archival 5
    minute HST/WFPC I exposure.

    From the FIRST (Faint Images of the Radio Sky at Twenty-Centimeters)
    log {\em N}--log {\em S}, we estimate that the probability of
    finding a background radio point source of the same or larger flux
    within 13 arcseconds from Circinus~X-1 is $\sim 1\%$.  In addition,
    given the IR source density in the region, the likelihood of
    finding an IR background source with equal or larger flux
    coincident with the radio and X-ray point source within the MIPS
    PSF is also of order 1\%.  Neither
    scenario can be ruled out statistically.  Thus, we cannot draw any
    conclusions about the nature of the SE point source without
    further observational evidence, e.g., from optical spectroscopy,
    variability, or proper motion.}
\end{itemize}

A detailed PSF subtraction of the binary emission (complicated by
effects of pileup) and spectral fits of these sources are beyond the
scope of this Letter and will be presented in a follow-up paper.

\subsection{Extended Diffuse Emission: X-ray Caps} 
\label{section:outer_emission}

The most obvious features in 
Figures~\ref{fig:outer_radio_overlay}-\ref{fig:csmooth} are the two
diffuse emission regions between $\sim 20 \arcsec$ and $50 \arcsec$ from Circinus
X-1 in the NW and SE directions.  The positions of both regions are
consistent with the diffuse emission tentatively reported in
\citet{heinz07b} and \cite{soleri09}.  Based on their morphological
appearance (NW cap appears slightly concave) and their placement away
from the binary along the jet axis, we will refer to both regions as
``caps'' throughout the rest of this Letter.  However, the geometry of
these regions could be complicated by projection effects, and other
interpretations are possible.

The caps have distinctly different X-ray colors from both the much
redder large-scale diffuse background (dominated by the dust-scattering
halo) and the distinctly bluer PSF, as can be seen from the
color image in Figure~\ref{fig:csmooth}.  The two caps have very
similar surface brightnesses ($\sim 2 \times 10^{-16}$ erg cm$^{-2}$
s$^{-1}$ arcsec$^{-2}$) and appear to be similar in their angular
extent (NW: $-5 \degree$--$65 \degree$; SE: $185 \degree$--245$\degree$).  However, they
are asymmetric in their radial extent (NW: $\sim 22 \arcsec$--50$\arcsec$, SE:
$\sim$18--45$\arcsec$).  The inner NW jet emission (Section~3.2) has similar
position and opening angles to the NW cap.

The correspondence between the high-resolution radio contours ($\ge 2
\sigma$) and the X-ray cap emission in
Figure~\ref{fig:inner_radio_overlay} is striking.  While the X-ray
image has significantly higher angular resolution than the radio, it
is clear that the X-ray caps align closely with the extended jet
emission.

The diffuse emission to the SE of the XRB in the radio image shows a
sharp surface brightness drop at the position of the cap and what has
been interpreted as a bend in the jet direction at this position,
trailing to the south (referred to as ``knot B'' in
\citealt{tudose06}).  We see no clear corresponding surface brightness
enhancement to this trail in the X-rays.  However, our sensitivity is
compromised by the fact that the read streak runs right along this
feature.

\begin{figure}[tbp]
  \centering
  \rotatebox{90}{\resizebox{!}{\columnwidth}{\includegraphics{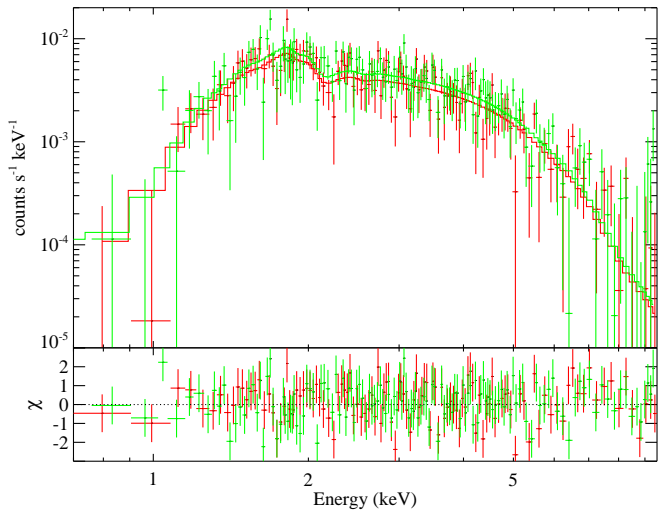}}}
  \hspace*{4em}
  \caption{Top: spectrum of the two large-scale caps and best-fit
    absorbed power-law fit (red: NW cap; green: SE cap).  Bottom:
    deviations from the best-fit model.
    \label{fig:spectrum}}
\end{figure}

We used partial annuli regions to extract the cap spectra.  Because
the background is dominated by the dust-scattering halo, the surface
brightness of which depends on the angular distance to \cir, we
carefully selected background partial annuli at the same radial extent
as the source regions.  We recover a total of 6240 source counts over
2860 expected background counts in both cap regions combined (0.5--9.5
keV).  The resulting spectra are shown in Figure~\ref{fig:spectrum}.

Initial extractions of each of the caps indicated that the spectral
parameters are statistically indistinguishable.  Therefore, we jointly
fit the spectra.  We restricted our spectral analysis to 0.5--9.5~keV.
Quoted uncertainties correspond to 90\% confidence intervals.  The
spectra are well fit by an absorbed powerlaw with a neutral hydrogen
column of $N_{\rm H}=2.24_{-0.28}^{+0.30} \times 10^{22}\,{\rm
  cm^{-2}}$ and a photon index, $\Gamma=1.99_{-0.21}^{+0.23}$ ($\chi^2
= 243.1$ over 252 degrees of freedom), giving 0.5--9.5 keV cap fluxes
of $3.2 \times 10^{-13}$ erg cm$^{-2}$ s$^{-1}$ (NW) and $2.7 \times
10^{-13}$ erg cm$^{-2}$ s$^{-1}$ (SE).

The caps can also be fit by a thermal (APEC) model ($\chi^2 = 237.7$
over 251 degrees of freedom).  The required absorbing column is very
similar to the power-law case with $N_{\rm H}=1.93_{-0.18}^{+0.21}
\times 10^{22}\,{\rm cm^{-2}}$.  This is consistent with the fact that
there is no obvious line emission present in the spectra, the APEC fit
requires a low abundance of $Z = 0.41_{-0.27}^{+0.38}$ and a high
temperature of $kT = 6.6_{-1.8}^{+2.2}$ keV.  When forced to solar
abundance, the fit requires even higher temperatures.  The 0.5--9.5
keV cap fluxes are $3.2 \times 10^{-13}$ erg cm$^{-2}$ s$^{-1}$ (NW)
and $2.7 \times 10^{-13}$ erg cm$^{-2}$ s$^{-1}$ (SE).

\section{Discussion} \label{section:discussion}

The limb-brightened morphology of the X-ray image (which has
significantly higher angular resolution than either radio image)
suggests that the emission from the two caps originates at a shock in
the outflow from \cir, seen in projection.  This suggestion is
supported by a sharp drop in surface brightness in radio and X-ray
just outside of the caps and by the fact that the SE radio jet seems
to be changing direction at the location of the X-ray cap.  The fact 
that the caps appear inside the large-scale radio nebula
would be a result of foreshortening, since the jet axis is likely
inclined with respect to the line of sight (though the actual
inclination is unknown).  The outflow itself appears largely X-ray 
dark (except for the inner NW diffuse radio--X-ray feature), similar 
to the X-ray cavities observed in many clusters with central radio 
galaxies.

The caps span projected half--opening angles of $35 \degree$ (NW) and
$30 \degree$ (SE).  This implies that either:
\begin{itemize}
\item[1.]{the outflow is, in fact, much wider than what would typically
    be considered a jet and might thus be better characterized as a
    non-thermal wind, or}
\item[2.]{the jets are precessing with a fairly wide opening angle, as
    is the case for SS433 (\citealt{margon84}), and/or}
\item[3.]{the jet or precession cone axis is very close to the line of
    sight, causing significant foreshortening of an intrinsically
    narrow jet.}
\end{itemize}
The wide opening angle of the {\em inner} NW jet would be consistent
with (1) and (3), but to be consistent with (2), it would require a
precession period of the jet that is short compared to the travel time
through the X-ray emission region to blend the jet emission \citep[as
seen in the arcsecond radio jets of SS433][]{hjellming81}.

The claim of a highly relativistic jet 
\citep[$\Gamma_{\rm jet} \sim 16$, $\theta \lesssim 5^{\circ}$;][]{fender04} 
coupled with these new observations would require that such a jet has a 
precession cone that is within $5 \degree$ of the line--of--sight such that 
the jet sometimes points very close to the line--of--sight.  Such a geometry 
would imply a physical scale of $\sim 20$~pc from cap to cap.  For comparison,
this is an order of magnitude larger than the projected distance
between the hotspots of XTE~J1550--564 \citep{corbel02} and H1743--322
\citep{corbel05} but would still fit comfortably within the radio
nebula surrounding SS433 \citep{dubner98}.

Below, we will discuss two possible interpretations of the radiative
origin of the cap emission.

\subsection{Cooled Synchrotron Model} \label{section:synch}

The most likely interpretation of the spectra is synchrotron emission.
In this case, the emission arises from the reaccelerated jet particles
entering the terminal shock, as is observed in the hot spots of many
FR~II radio galaxies \citep[e.g.,][]{meisenheimer89}.  The X-ray
emission in this case should be co-spatial with the radio, which
appears consistent with the radio contours in
Figure~\ref{fig:inner_radio_overlay}.

The radio synchrotron spectrum of the large-scale nebula has a typical
spectral index of $\alpha_{\rm R}=0.53$ (with $F_\nu \propto \nu^{-
  \alpha_{\rm R}})$ \citep{tudose06}, consistent with the standard
power-law slope of first-order Fermi acceleration.  For the canonical
model of continuously injected power-law electrons one would expect a
spectral index of $\alpha_{\rm X}=\alpha_{\rm R} + 0.53 = 1.03$ above
the cooling break at frequency $\nu_{\rm b}$, perfectly consistent
with the observed X-ray spectral index of $\alpha_{\rm X}=0.99$.  From
extrapolation of the radio and X-ray spectra, we estimate the break
frequency from uncooled to cooled synchrotron emission to be
$\nu_{\rm b} \sim 2\times 10^{16}$ Hz, which is uncertain by about an
order of magnitude when uncertainties in the radio and X-ray spectral 
indices are used primarily because the X-ray and radio power-law
slopes are extrapolated over a very large range of frequencies.

Taking the emission regions to be spherical to lowest order, we
estimate the equipartition magnetic field strength to be $B_{\rm eq}
\sim 50 \, \mu$G, which gives a {\em minimum} total internal energy
for each emission region of $\sim 8 \times 10^{45}$ erg. From the 
estimate of the break frequency, we estimate a cooling time, 
$\tau_{\rm cool} \sim 1600 \,{\rm yr}$.    A different emission geometry 
will change these numbers by factors of order unity.

Together, these numbers provide a robust {\em lower} limit on the
total jet power of $P_{\rm cir} > 3 \times 10^{35}\,{\rm
  erg\,s^{-1}}$, needed {\em just} to put the synchrotron emitting
particles in place (not including any plasma currently in the jet or
the energy needed to inflate the large-scale radio nebula).  Departing
from equipartition can only {\em increase} this number.  This limit
is {\em independent} of the assumed geometry and size of the emission
regions.

Conversely, we can place a rough {\em upper} limit on the outflow
power from the fact that the large-scale radio nebula must be at least
as old as $\tau_{\rm cool}$.  Using the same arguments as presented in
\cite{heinz02} and \cite{tudose06} for the expansion of the radio
nebula we find that $P_{\rm cir} \lesssim 2 \times 10^{37}\,{\rm
  erg\,s^{-1}}$, though this limit is not as robust since the magnetic
field could well be out of equipartition and the nebula could be
significantly elongated along the line--of--sight, both of which would
increase the total power.

These limits on $P_{\rm cir}$ confirm that the accreting compact
object in this system is driving powerful jets into the interstellar
medium (ISM).  Taking the high 42 year average luminosity of Circinus~X-1
at about 80\% of the Eddington luminosity for a 1.4 $M_{\odot}$ neutron
star \citep{parkinson03} at face value would imply that the long-term
average jet power is between 2\% and 10\% of the {\em current} average
radiative power.

Given that the average long-term accretion rate cannot greatly exceed
the Eddington rate, we can derive a robust lower limit on the
efficiency with which accretion power is converted into outflow power
of $\eta_{\rm jet} = P_{\rm cir}/\dot{m}c^2 > 0.034\%\frac{\dot{m}_{\rm
    Edd}}{\langle \dot{m}\rangle}$ (see also \citealt{heinz07b} and
\citealt{soleri09}).

\subsection{Thermal Model} \label{section:thermal} 

Given the equally good fit we achieved with a thermal emission model,
it is worth discussing the alternative scenario that the X-ray caps
are the shocked ISM pushed ahead of the expanding radio outflow
(similar to the shells around many X-ray cavities in galaxy clusters
and around the microquasar Cygnus~X-1; \citealt{gallo05}).

In the thermal model the X-rays should be located further out from the
binary than the radio emission.  However, the high-resolution radio
data indicate that the radio and the X-ray emission are co-spatial.
This requires a very low inclination angle in order to align the two
regions by projection \citep[which is consistent with the inclination
required by the ultra-relativistic flow claimed in][]{fender04}.

From the best-fit temperature of 6.6 keV, we directly infer a shock
velocity of $\sim 2.4 \times 10^3$ km s$^{-1}$.  From the projected
distance, we can infer an estimated travel time of $\sim
500 / {\rm sin}(\theta)$ years out to the observed shock positions for each
of the caps.  Again approximating the emission region as spherical, we
estimate the mass and the total (thermal plus kinetic) energy of the
shocked material for each of the caps to be $\sim 2 \times 10^{32}$ g
and $\sim 2 \times 10^{49}$ erg, respectively.  Combined with the
shock travel age, this gives a minimum outflow power for both caps
combined of $P_{\rm cir} > 6 \times 10^{38}\,\times\,\sin{(\theta)}$
erg$^{-1}$ s$^{-1}$, {\em just} to supply the X-ray emitting material (not including
the putative power required to inflate the large-scale radio nebula).
Unless the inclination angle is very small, this power is exceedingly
large.  Even for inclinations as low as those inferred from the
relativistic jet claimed in \cite{fender04}, the {\em minimum} average
power would still be $P_{\rm cir} > 5\times 10^{37}\,{\rm erg\,s^{-1}}$.

While we cannot completely rule out a thermal origin, the very high power 
required by the thermal model combined with the perfect spectral 
correspondence to a classic broken synchrotron spectrum and the spatial 
correspondence with the radio suggest that the synchrotron model is the 
more natural choice.

\section{Summary}

We have presented an initial analysis of a deep {\em Chandra} imaging
observation of \cir.  We detect two diffuse X-ray caps that are likely
the terminal shocks of powerful jets running into the ISM.  In
addition, we find an arcsecond-scale outflow between the XRB and one
of the X-ray caps, coincident with the radio jet. These discoveries
make Circinus~X-1 the first microquasar with both an X-ray jet
and two stationary X-ray shocks, and one of only a handful of
microquasars with a direct estimate of the jet power.

\acknowledgments {S.H. and P.S. acknowledge support from NASA grant
  GO9-0056X.}


\end{document}